\documentclass[aps,prX,twocolumn,reprint,superscriptaddress,longbibliography]{revtex4-1}

\usepackage[final]{graphicx}
\usepackage{dcolumn}
\usepackage{multirow}
\usepackage{bm}
\usepackage{floatflt,epsfig}
\usepackage{float}
\usepackage{amsmath}
\usepackage{xfrac}
\usepackage{amsfonts}
\usepackage{amssymb}
\usepackage{mathrsfs}
\usepackage[latin1]{inputenc}
\usepackage{siunitx}
\DeclareTextSymbol{\degre}{T1}{6}
\DeclareTextSymbol{\degre}{OT1}{23}

\begin{document}

\title{Ultrafast Surface Plasmonic Switch in Non-Plasmonic Metals}

\author{E. B\'{e}villon}
\affiliation{Laboratoire Hubert Curien, UMR CNRS 5516, Universit\'{e} de Lyon,\\ Universit\'{e} Jean-Monnet, 42000
Saint-Etienne, France}
\author{J.P. Colombier}
\email{jean.philippe.colombier@univ-st-etienne.fr}
\affiliation{Laboratoire Hubert Curien, UMR CNRS 5516, Universit\'{e} de Lyon,\\ Universit\'{e} Jean-Monnet, 42000
Saint-Etienne, France}
\author{V. Recoules}
\affiliation{CEA-DIF, 91297 Arpajon, France}
\author{H. Zhang}
\affiliation{Laboratoire Hubert Curien, UMR CNRS 5516, Universit\'{e} de Lyon,\\ Universit\'{e} Jean-Monnet, 42000
Saint-Etienne, France}
\author{C. Li}
\affiliation{Laboratoire Hubert Curien, UMR CNRS 5516, Universit\'{e} de Lyon,\\ Universit\'{e} Jean-Monnet, 42000
Saint-Etienne, France}
\author{R. Stoian}
\affiliation{Laboratoire Hubert Curien, UMR CNRS 5516, Universit\'{e} de Lyon,\\ Universit\'{e} Jean-Monnet, 42000
Saint-Etienne, France}
\date{\today}

\begin{abstract}
We demonstrate that ultrafast carrier excitation can drastically affect electronic structures and induce brief surface  
plasmonic response in non-plasmonic metals, potentially creating a plasmonic switch. Using first-principles molecular 
dynamics and Kubo-Greenwood formalism for laser-excited tungsten we show that carrier heating mobilizes $d$ electrons 
into collective inter and intraband transitions leading to a sign flip in the imaginary optical conductivity, 
activating plasmonic properties for the initial non-plasmonic phase. The drive for the optical evolution can be 
visualized as an increasingly damped quasi-resonance at visible frequencies for pumping carriers across a chemical 
potential located in a $d$-band pseudo-gap with energy-dependent degree of occupation. The subsequent evolution of 
optical indices for the excited material is confirmed by time-resolved ultrafast ellipsometry. The large optical 
tunability extends the existence spectral domain of surface plasmons in ranges typically claimed in laser self-organized 
nanostructuring. Non-equilibrium heating is thus a strong factor for engineering optical control of evanescent 
excitation waves, particularly important in laser nanostructuring strategies.
\end{abstract}


\maketitle

\section{Introduction}

Collective non-equilibrium effects on transport properties in metallic systems are topics of current interest as
non-standard behaviors play an increasing role in optics, thermodynamics or magnetism. Ultrafast laser electronic
heating can mobilize localized states, inducing massive nonlinear contributions to optical and thermal transport or
structural metastability. Understanding the intimate electronic mechanisms of optical coupling in excited solids and
controlling excitation transients is therefore of prime importance as new applications are emerging in ultrafast optics,
plasmonics and heat control strategies. If consequences for transport are readily expectable \cite{Desjarlais02,
Chen13}, and the role of nonequilibrium electronic transfer was early recognized \cite{Elsayed-Ali87, Schoenlein87},
subtle atomistic effects can be inferred at the level of the band structure with unexpectedly strong consequences.
Charge transfer and screening during non-equilibrium concur to a dynamic self-adjustment of thermal and optical
properties to accommodate swift excitation. Recoules \textit{et al.} \cite{Recoules06} proved that participation of
localized $d$ orbitals in noble metals enforces the lattice cohesion. Non-equilibrium charge supply allows large
variations of thermal characteristics in transition metals \cite{Lin08}, fluidifying energy transport \cite{Petrov13}.
Non-thermal distributions can restrict the collisional phase-space around the Fermi level, severely damping
electron-phonon coupling \cite{Mueller13}. Furthermore, electronic occupation of delocalized states and filling from
localized reservoirs (e.g. $d$-bands in transition metals) redefine classical views on free electron density
\cite{Bevillon14a}, introducing significant dynamics in optical behaviors. Tuning ultrafast optical response via
electronic reactions was thus proposed for active large bandwidth modulation in ultrafast plasmonics
\cite{MacDonald09}. Engineering macroscopic optical response on ultrafast scales equally relates to the onset of optical
resonances \cite{Lukyanchuk10, Hessel65} on excited surfaces with spectral and spatial disturbances, e.g. Wood's
anomalies. One debated example concerns ultrafast laser nanostructuring of solids in self-assembled regular patterns.
The universal phenomenon - observed half a century ago \cite{Birnbaum65} - carries application potential in functional
surfaces \cite{Zorba08}, color-coded optical traceability and multidimensional information storage
\cite{Dusser10,Zhang14}, or feedback-driven nanolithography \cite{Oktem13}. Key questions in laser-induced periodic
surface structures (LIPSS) concern the origin of spatially-modulated energy patterns \cite{Sipe83} and the interplay
between optical resonances and nonlinear feedback. Electromagnetic calculations indicate the involvement of surface
waves inducing collective motion \cite{Sipe83}. Among them, surface plasmon (SP) with its ability to enhance
electromagnetic energy on nanoscales holds a determinant role.

The potential involvement of SPs interference in laser nanostructuring is in close relation to surface optical indices.
Thus the experimentally-observed onset of polarization-dependent periodic structures for non-plasmonic metals, as
demonstrated on tungsten \cite{Vorobyev08}, is intriguing in optical ranges theoretically forbidden by predictions based
on ambient optical indices. A typical example of ultrafast LIPSS at 800\,\si{nm} is provided in
Fig.~\ref{ti-te-evolution}(a), showing quasi-wavelength periodicity. The positive real permittivity of W precludes
normally the generation of electronic surface waves and plasmon polariton coupling. The possibility to excite collective
material motion pinpoints to a more intricated dynamic response of the system. We investigate the impact of ultrafast
heating on electronic structures capable of initiating a large excursion of optical properties, namely an
excitation-driven plasmonic state in otherwise non-plasmonic solid non-Drude-like metal. First-principles approaches are
chosen to interrogate electronic-driven evolution of optical properties and their consequences on plasmonic behaviors at
time and spatial scales difficultly accessible by experiments.

\section{Calculation details}

The calculations correspond to ultrashort pulse irradiation conditions enabling LIPSS around the damage threshold
(90\,\si{mJ cm^{-2}} absorbed fluence). Transient electron and lattice temperatures in W for single ultrashort pulse
irradiation (Fig.~\ref{ti-te-evolution}(b)) are estimated using a two-temperature hydrodynamic code (Esther)
\cite{Colombier12b} which describes the energy balance using Helmholtz optical formalism and electron-phonon relaxation
with temperature-dependent transport properties \cite{Lin08, Bevillon14a}. Electronic temperatures on the rising heat
cycle can amount to $2.5 \times 10^{4}$\,\si{K}, inducing electron-phonon nonequilibrium up to 3\,\si{ps}, with the
material in solid state for at least 1\,\si{ps}. Internal electronic thermalization is assumed, justified by
high levels of energy deposition in the vicinity of the threshold which accelerates electronic energy exchange to fs
scales \cite{Mueller13}. In this range, calculations are carried out with the plane-wave code Abinit \cite{Gonze09}, in  
the frame of the density functional theory \cite{Hohenberg64, Kohn65} extended to finite electronic temperatures 
\cite{Mermin65}. The generalized gradient approximation \cite{Perdew96} is employed to model exchange and correlation 
energies, and projector augmented-waves atomic data \cite{Torrent08} account for the effects of nuclei and core 
electrons. A 54 atoms supercell of body centered cubic (bcc) W is considered, with 5$s^{2}$5$p^{6}$5$d^{4}$6$s^{2}$ 
valence electronic configuration; 4$f$ electrons are neglected as their effect is weak at the $T_e$ considered here 
\cite{Bevillon14a}. We first perform \textit{ab initio} molecular dynamics simulations in the isokinetic ensemble at 
room temperature, with $T_i = T_e = 300$\,\si{K} during 2\,\si{\pico\second}. This calculation provides an average 
thermodynamic equilibrium from which ionic configurations are extracted, standing for representative states of the W 
lattice at ambient conditions. Then electronic structures are computed at $T_e$ of $300$, $10^{4}$ and $2.5 \times 
10^{4}$\,\si{K}. They serve as a basis to evaluate optical properties in solid phase at these levels of electronic 
excitation.

\section{Results}

\begin{figure}[ht]
\includegraphics[width=8.5cm]{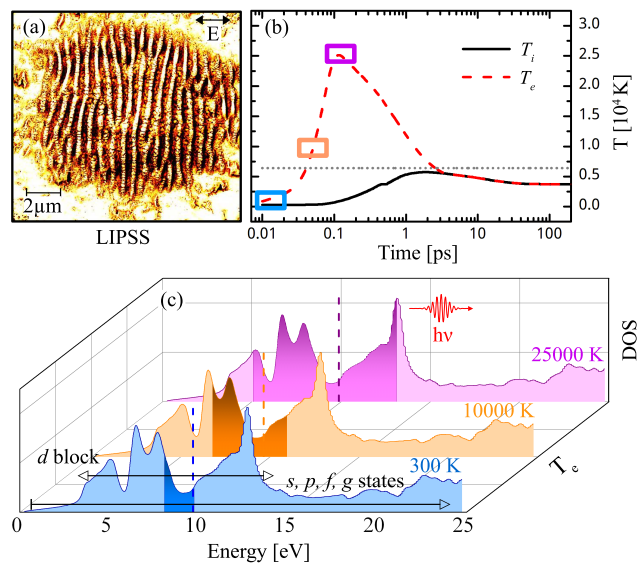}
\caption{\small (Color online) (a) Typical field-perpendicular surface periodic nanostructures at quasi-wavelength
periodicity induced by 800\,\si{nm} 50\,\si{fs} laser pulses on W. (b) Electron (dashed red) and lattice (solid black)
temperature evolutions for W irradiated by a 50\,\si{fs} laser pulse. Dotted line indicates the standard vaporization
limit. (c) DOS of bcc W computed at $T_i = 300$\,\si{K} and its evolution with $T_e$. Dashed lines indicate the Fermi
levels and dark colored areas show occupied electronic states mainly impacted by 800\,\si{nm} photons.}
\label{ti-te-evolution}
\end{figure}

We first focus on the evolution of the electronic density of states (DOS) with electronic temperatures in the presence of a
cold, ambient lattice, with calculated DOS profiles provided in Fig.~\ref{ti-te-evolution}(c). If finite ionic
temperature induces a certain smoothing of the features compared to calculations at $T_i$ = 0\,\si{K}
\cite{Bevillon14a} due to a loss of degeneracy arising from atom oscillations around their high symmetry position, a
highly-structured $d$ band remains. The DOS shape is remarkably stable against electronic heating with a maximal $d$
block shift of 0.15\,\si{eV} at $2.5 \times 10^{4}$\,\si{K}, indicative of a stiff electronic structure. This relates to
the roughly symmetric profile of the DOS on both sides of the electronic chemical potential $\mu$. The high density of
empty electronic states on the right side of the Fermi level collects the excited electrons from the left-sided filled
electronic states, leading to a weak $T_e$ dependence of $\mu$, contrary to most of transition metals \cite{Lin08,
Bevillon14a}. The location of the chemical potential within a mid-range pseudo-gap between high density of filled and
empty electronic states determines the non-plasmonic behavior in the 1-5\,\si{eV} range, i.e. for visible photon
energies bridging the gap. The increase of $T_e$ determines however a slight augmentation of the number of $d$ electrons
from low-lying part of $sp$ orbitals, implying a moderately strengthen localization of the charge density, noted in
increasing Hartree energies \cite{Bevillon14a}. The main $T_e$ effect resumes to a Fermi broadening within the DOS. The
impact of the occupation probability is discussed below.

Relying on the as determined electronic structures, the subsequent optical properties are obtained from an average
on three ionic configurations. The real part of the frequency-dependent optical conductivity is obtained within the
Kubo-Greenwood (KG) formalism \cite{Mazevet10}:

\begin{eqnarray}
\displaystyle \sigma_r(\omega) = \frac {2\pi} {3\omega\Omega} \sum_{i,j,k} \sum_{\alpha=1}^{3} [f(\epsilon_{i,k}) -
f(\epsilon_{j,k})] \nonumber\\
\times | \langle \Psi_{j,k} | \nabla_\alpha | \Psi_{i,k} \rangle |^2 \delta(\epsilon_{j,k}-\epsilon_{i,k}
-\hbar\omega), \label{KG}
\end{eqnarray}

\noindent where electronic transitions from $i$ to $j$ states are integrated over the reciprocal space for each photon
energy $\hbar\omega$, accounting for the Fermi-Dirac occupations $f(\epsilon)$ and the eigenenergies of the electronic
states $\epsilon$. $\nabla$ represents the velocity operator. $\Omega$ is the volume of the cell while $\alpha$
corresponds to the three spatial dimensions. The imaginary part of the frequency-dependent conductivity is obtained
using the Kramers-Kronig (KK) relation, $ \sigma_i(\omega) = - \frac {2} {\pi} {\cal{P}} \int_{0}^{\infty} \frac
{\sigma_r(\omega') \omega} {\omega'^2 - \omega^2} d\omega'$, where $\cal{P}$ is the principal value of the integral.
Frequency-dependent permittivities and optical indices can be derived.

\begin{figure}[ht]
\includegraphics[width=8.5cm]{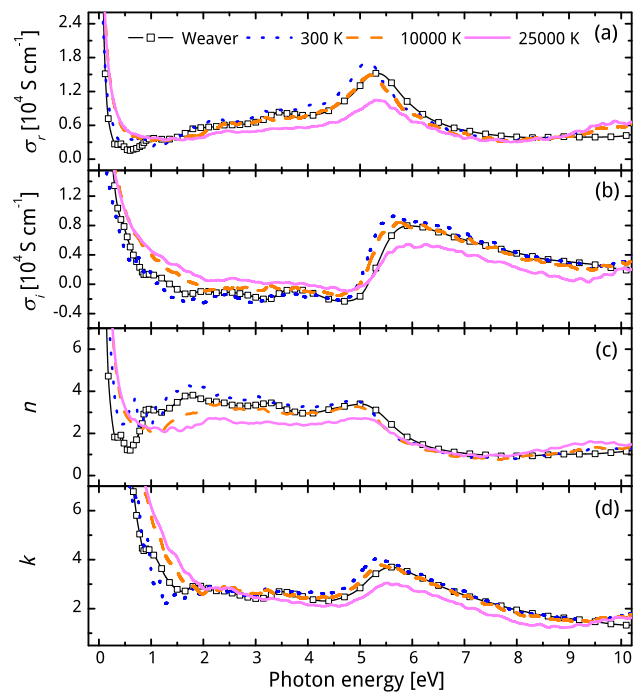}
\caption{\small (Color online) Real (a) and imaginary (b) part of the frequency-dependent optical conductivity; real
(c) and imaginary (d) part of the optical indices. Colored curves stand for the electronic temperatures of
$300$\,\si{K}, $10^{4}$\,\si{K} and $2.5 \times 10^{4}$\,\si{K} and the squares represent the experimental data for
non-excited W \cite{Weaver75}.}
\label{optical}
\end{figure}

Optical properties follow electronic structure evolutions and large excursions up to plasmonic states are argued below. Fig.
~\ref{optical} shows the computed frequency-dependent optical conductivities $\tilde{\sigma} = \sigma_r + i\sigma_i$ and
optical indices $\tilde{n} = n + ik$ as a function of the electronic temperature. A good agreement is found between
theoretical values obtained at $300$\,\si{K} and the reflectivity measurements of Weaver \cite{Weaver75}, confirming a
realistic description of the W electronic structure. $\sigma_r$ is an implicit measure of optical absorption and its
spectral behavior can be inferred based on the profile of the energy bands calculated in Fig.~\ref{ti-te-evolution}(c).
For the low $T_e$ case (dotted blue line), at photon energies below 0.5\,\si{eV}, the intraband part dominates,
resulting from electronic transitions inside 6$sp$ bands. This contribution rapidly decays as the photon energy
increases. On the other hand, interband transitions originating from the partially filled $5d$ sub-bands become
gradually more important \cite{Romaniello06}. From 0.5 to 5.1\,\si{eV}, photon-driven electronic transitions access an
increasingly larger DOS domain ($\pm h\nu $) centered on $\mu$, resulting in a stepwise build-up of $\sigma_r(\omega)$
mapping the local DOS. This trend can be seen as the wing slope of a resonant behavior around the 5.1\,\si{eV} peak
given by the particular electronic DOS splitting around the Fermi level and the finite width of the sidebands. The
conductivity finally decreases once main $d$-bands peaks have been included and diluted into the continuously
increasing electronic transition phase-space. From the KK relation, $\sigma_i(\omega)$ relies on the overall
$\sigma_r(\omega)$ profile. Conceptually, this integral can be separated in two parts depending on the value of
$\omega'$ with respect to the reference value $\omega$. For $\omega' \in [0, \omega[$, the integral is negative and
reverse signs for higher $\omega'$. Accordingly, the magnitude of $\sigma_i$ depends on the profile of the real
conductivity mainly around $\omega$. A quasi-symmetric profile of $\sigma_r$ tends to balance positive and negative
components of the integral and provide low values for $\sigma_i$. On the contrary, an asymmetric profile emphasizes the
respective sign components of the integral, depending on the trend of asymmetry (left or right-turned), determining
positive or negative values of the imaginary conductivity. For W, the positive slope of $\sigma_r$ profile from 1 to
5.1\,\si{eV} with values ranging from $0.5 \times 10^{4}$ to $1.7 \times 10^{4}$\,\si{S cm^{-1}} leads to a negative
$\sigma_i(\omega)$ at ambient conditions. A spectral view on $\sigma_i(\omega)$ indicates an anomalous-like dispersive
behavior related to the absorption resonances.

The profiles of optical conductivities in Fig.~\ref{optical}(a,b) and optical indices in Fig.~\ref{optical}(c,d)
significantly change with the electronic temperature, requiring an introspection in the corresponding excitation-driven
electronic effects. If the observed changes in optical properties are not a consequence of DOS variation due to its
stability against heating, the electronic temperature affects the degree of filling. Thus the optical evolution is the
direct consequence of a broaden Fermi-Dirac charge redistribution, making more states available for photon-induced
transitions. With the increase of $T_e$, the optical transitions concern an increasing $\mu$-centered interval mainly
corresponding to [$-h \nu - \frac {3}{2}k T_e$, $h \nu +\frac {3}{2}k T_e$] \cite{Hopkins08}, illustrated in
Fig.~\ref{ti-te-evolution}(c) for a photon energy of 1.55\,\si{eV}. The consequence is manyfold, observable in the
$\sigma_r$ which maps the spectral absorption. Firstly this produces a broadening of available transition
phase-space for the various photon energies. This translates into an increase of the intraband part, balanced by a
decrease in the interband domain (qualitatively similar to a clockwise turn), damping the $\sigma_r$ 5.1\,\si{eV} peak.
The subsequent asymmetric change of slope in $\sigma_r$ with $T_e$ determines the increase of $\sigma_i$ in the
1-5\,\si{eV} (as the dispersive behavior across the resonance flattens) and a passage in the positive values domain for
a significant part of the spectral domain. The situation becomes more clear in the evolution of the optical indices. At
1.55\,\si{eV} (800\,\si{nm}), relevant for the LIPSS, optical indices are strongly affected by the heating of the
electronic system, with $\tilde{n} = 4.0+2.4 i$ at $3 \times 10^{2}$\,\si{K} reaching $\tilde{n} = 2.1+3.8 i$ at $2.5
\times 10^{4}$\,\si{K}. Thus, in the near infrared and low frequency visible part of the spectrum, the real part of the
index goes down and the imaginary part augments.

\section{Experimental confirmation}

The evolution was confirmed by time-resolved pump-probe ultrafast ellipsometry on laser-excited W surfaces. The  
transient optical properties upon ultrafast (120\,fs) laser irradiation were probed at 1.55\,eV photon energy using a 
two-angle one color time-resolved ellipsometry method following the technique proposed in Ref. \cite{Roes03}. The static 
properties of the non-excited surface were first evaluated ex-situ using a commercial ellipsometer (Uvisel, Horiba 
Jobin Yvon) and the results give $\tilde n=3.57 + i3.15$ for massive W materials (Goodfellow $99.95\%$ purity, 
mechanically polished). Alongside massive W, foils (Goodfellow 0.3\,mm thick, polished at 0.1\,mRa -purity $99.95\%$) 
were equally used as reference. The dynamic reflectivity changes were subsequently interrogated on massive thick targets 
by p-polarized 120\,fs, 800\,nm low energy non-perturbing probe laser pulses at 27.1\,$^{\circ}$ and 65.8\,$^{\circ}$ 
incidence angles. The probe pulses were time-synchronized with fs accuracy with the exciting pump pulse of equally 
120\,fs 800\,nm, arriving at normal incidence on the W surface. Two photodiode detectors were used in imaging 
geometries with respect to the surface. The probed zone is significantly smaller than the spatial extent of the excited 
region. The exciting fluences were chosen slightly below the ablation threshold.

\begin{figure}[h]
\includegraphics[width=8.5cm]{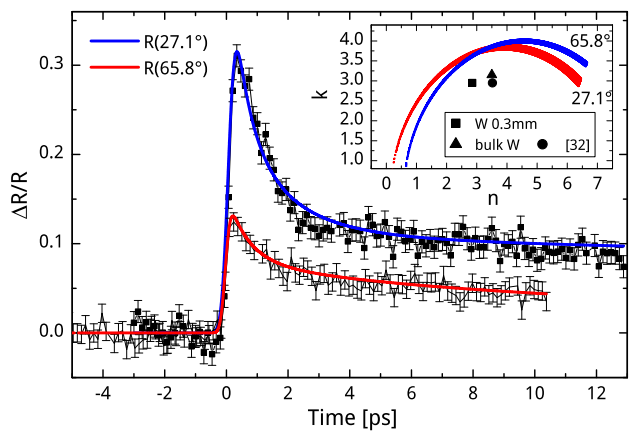}
\caption{\small (Color online) Example of dynamic time-resolved reflectivity traces at the two probe angles 
(27.1\,$^{\circ}$ and 65.8\,$^{\circ}$) at incident pump peak fluence of 0.12\,J/cm$^{2}$. Inset: ($n,k$) contour plots 
corresponding to measured reflectivities at the two given angles. The intersection point represents a uniquely 
determined ($n,k$) pair satisfying simultaneously the two reflectivity conditions. Non-excited values obtained from 
ellipsometric measurements on bulk and foil samples as well as literature
data \cite{Rakic98} are given as references.}
\label{fres}
\end{figure}

The swift optical activity on massive W samples is underlined by strong reflectivity changes, with a reflectivity  
snapshot example for an input peak fluence of 0.12\,J/cm$^{2}$, just below damage threshold being shown in the 
Fig.~\ref{fres} (a). The time-profile of the reflectivity transients maps the heating/cooling cycle of the electronic 
system and the associated redistribution in the electronic occupation around the chemical potential with $T_e$. The 
measured maximum values of the transient reflectivity changes occurring just after the peak of the excitation pulse 
were then depicted in terms of corresponding ($n,k$) values in the ($n,k$) optical phase-space. They represent 
uniquely-determined ($n,k$) pairs obtained by inverting Fresnel formulas at the given angles. The result is presented in 
Fig.~\ref{fres} (b). The intersection point of the ($n,k$) contour plots corresponding to two angle-resolved measured 
reflectivities were used to extract the corresponding ($n,k$) pair, allowing thus access to the  evolution of optical 
indices. The result is $\tilde n = 3.21 + i3.77$, where resulting optical indices indicate slight $n$ decrease and 
increase in $k$ above the experimental value of its dispersive part. Note that these correspond to a macroscopic state 
averaged over a region set by the optical penetration depth of 18\,nm, with inhomogeneous temperature distribution. The 
accuracy of the measurement is affected by the roughness and local planarity of the surface and care was taken to 
minimize the errors. The optical evolution has strong impact on the possibility to excite surface plasmon, with the 
fulfilment of the required optical conditions.

\begin{figure}[ht]
\includegraphics[width=8.5cm]{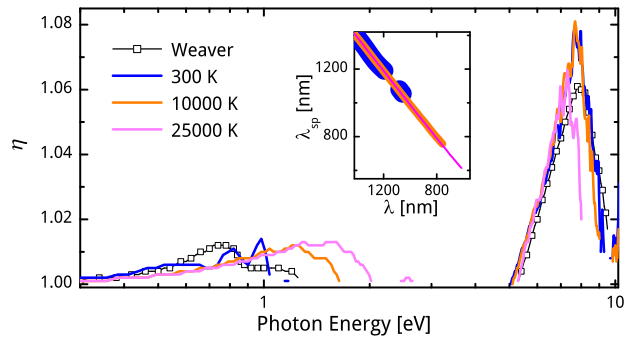}
\caption{\small (Color online) Real part of the effective refractive indices ($\eta$) in case of air-tungsten interface
as a function of $T_e$. The symbol line stands for experimental data. Inset: variation of plasmon wavelength as a
function of laser wavelength in a spectral domain ranging from 0.9 to 2.3\,\si{eV}.}
\label{eri}
\end{figure}

For air-material interfaces the condition for surface plasmon existence reduces to $\varepsilon_r = n^2 - k^2 < -1$, a  
condition fulfilled in the conditions of Fig.~\ref{fres}, marking thus the ultrafast plasmonic activation at fluencies 
in the close vicinity of the damage threshold.

\section{Discussion}

Surface plasmon periodicity $\lambda_{sp}$ as a function of laser wavelength $\lambda$ is given by $\lambda_{sp} =
\lambda /\eta$, where $\eta = \Re\{[\tilde{n}^2/ (\tilde{n}^2+1)]^{1/2}\}$ is the real part of the effective refractive
index \cite{Pitarke07}. Calculated $\eta$ is plotted in Fig.~\ref{eri} and the absence of data indicates non-existence 
domains of surface plasmons. With the increase of $T_e$, one can note the expansion of the existence domain in the red 
photon energy region. The temperature-induced broadening of the electronic transition domain determines thus a plasmonic
switch at visible optical frequencies. It appears that, upon electronic excitation, the light-induced onset of
plasmonic character can sustain an origin based on optical resonances for LIPSS at 800\,\si{nm} even though room
temperature optical indices indicate non-plasmonic properties. As expected for air-metal interfaces, $\eta$ remains
close to one, especially at low photon energy, leading to $\lambda_{sp}$ slightly inferior to the laser wavelength
(inset of Fig.~\ref{eri}).

\begin{figure}[ht]
\includegraphics[width=8.5cm]{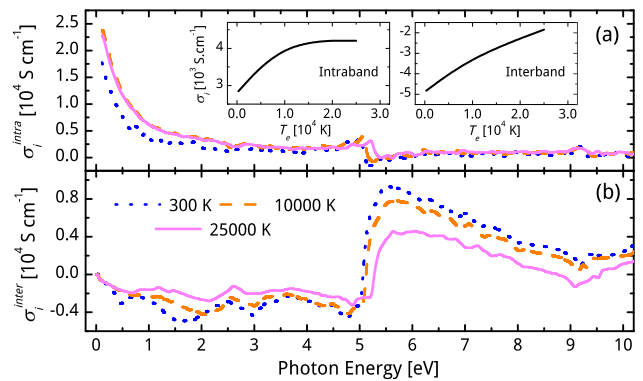}
\caption{\small (Color online) Intra and interband contribution to the imaginary conductivity as a function of $T_e$.
Inset: Local evolution of $\sigma_{i} (T_e)$ at 1.55\,\si{eV}.}
\label{interband}
\end{figure}

From the expression of the dielectric permittivity $\tilde\varepsilon =1+i\tilde\sigma/ \omega\varepsilon_{0}$ the
existence condition can be expressed more directly in terms of imaginary conductivity (and implicitly on real
permittivity) with $\sigma_i > 2 \omega \varepsilon_0 $. To clarify processes leading to the increase of the existence
domain, we propose a simplifying approach. The imaginary conductivity is tentatively split into intra and interband
components as depicted in Fig.~\ref{interband}(a,b). We first compute optical properties for a degenerate electronic
system at $T_i = 0$\,\si{K} and $T_e =$ $300$, $10^{4}$ and $2.5 \times 10^{4}$\,\si{K}. This disregards ionic
temperature and phonon effects, assuming they assist mostly the intraband component via momentum conservation
conditions. Conceptually this is justified in a classical view by the temperature dependence of damping frequencies.
Accordingly, the intraband contribution vanishes and $\sigma_{i}^{inter} \simeq \sigma_{i} (T_i=0K)$
(Fig.~\ref{interband}(b)). The intraband part at $T_i = 300$\,\si{K} is then extracted by subtracting the as-determined
$T_i$-insensitive interband part from the total value of the imaginary conductivity (Fig.~\ref{optical}(b)). In the
visible spectral range, the intraband part $\sigma_{i}^{intra}$ rapidly increases with the rise of $T_e$ and saturates
at $10^{4}$\,\si{K}. An example of this behavior at 1.55\,\si{eV} is given in the inset. At the opposite, the interband
part shows a more complex behavior, with negative values of $\sigma_{i}^{inter}$ in the 0 to 5\,\si{eV} energy interval
that leads to negative values of the total imaginary conductivity, explaining the non-plasmonic nature of this metal.
At higher energies, a resonance-like dispersive shape induced by the 5.1\,\si{eV} peak of $\sigma_{r}$ is clearly
visible (Fig.~\ref{interband}(b)). By weakening the peak of $\sigma_{r}$, Fermi broadening also flattens the shape of
$\sigma_{i}^{inter}$ around the resonance and gradually reduces its negative component, progressively switching on
plasmonic properties of W. The increase is not yet saturated at $2.5 \times 10^{4}$\,\si{K} giving the possibility of a
stronger lift and a subsequent extension of the plasmon existence domain for higher electronic temperatures. This
highlights the preponderant role of $d$-electron-driven interband transitions in the optical response at high $T_e$.

The electronic evolution described above is related to the fast achievement of the non-equilibrium electron-lattice phase. However, W is a strong coupling material and lattice effects may be rapidly generating along the electronic relaxation. Performing simulations at different lattice temperature values we observe that a ionic temperature effect is indeed observable, increasing slightly the plasmon existence domain. This is attributed to a loosening of the atomic order initially responsible for the material non-plasmonic nature. However the effect is of secondary importance compared to the charge redistribution induced by the electronic temperature within the band structure. We conclude that lattice heating, albeit its influence on atomic order, does not severely impact the calculated electronic structure on the relevant timescales, maintaining a dominant electronic drive for the process.

\section{Conclusion}

In conclusion, electron temperature dependent \textit{ab initio} MD-KG calculations of solid state W indicate large duty-cycle
optical tuning upon ultrafast laser irradiation and potential excitation of propagating collective electronic motion for an initially non-plasmonic state. We demonstrate that the evolution of optical properties with $T_e$ produces, via the necessary optical conditions, an extension of the predictable existence domain of surface plasmon in the visible range, rendering possible a transient plasmonic phase and a potential plasmonic involvement in LIPSS. The necessary condition for optical indices was confirmed by time-resolved ellipsometry on excited W surfaces. The dynamic evolution mechanism is related to a redistribution of localized $d$-electrons across a chemical potential located in a $d$-band pseudo-gap. Since the DOS is not distorted by electronic heating, the changes of plasmon properties, indicated here by
corresponding changes of optical conductivities, are principally due to Fermi broadening within a structured $d$-block, extending
the transition space for visible frequencies. By depopulating low-lying electronic states in favor of high electronic
states, quasi-resonant transitions from regions of high DOS are diluted in a continuum of transitions. The dissimilar
behaviors of intra and interband absorption events mark thus the evolution of the optical conductivity towards
fulfilling resonant conditions. Similar phenomena may occur in other
metals exhibiting non-plasmonic characteristics \cite{Rakic98}, especially Cr and Mo, as they crystallize in a similar
structure.  Stronger changes of optical properties \cite{Garrelie11} are to be expected in case of transition and noble
metals where the electronic structure varies substantially with electronic temperature. All these effects have
implications for optical tunability and ultrafast switching of plasmonic properties extending beyond the domain of laser
nanostructuring and validate concepts of structure engineering for metallic materials.

\section{acknowledgments}

We thank N. Faure and M. Torrent for experimental and computing support. This work was supported by ANR project DYLIPSS
(ANR-12-IS04-0002-01) and by LABEX MANUTECH-SISE (ANR-10-LABX-0075) of the Universit\'{e} de Lyon, within the ANR
program "Investissements d'Avenir" (ANR-11-IDEX-0007). Calculations used resources from GENCI (project gen7041).

\bibliographystyle{apsrev4-1-etal}
\bibliography{tungsten-v1}

\begin{thebibliography}{34}%
\makeatletter
\providecommand \@ifxundefined [1]{%
 \@ifx{#1\undefined}
}%
\providecommand \@ifnum [1]{%
 \ifnum #1\expandafter \@firstoftwo
 \else \expandafter \@secondoftwo
 \fi
}%
\providecommand \@ifx [1]{%
 \ifx #1\expandafter \@firstoftwo
 \else \expandafter \@secondoftwo
 \fi
}%
\providecommand \natexlab [1]{#1}%
\providecommand \enquote  [1]{``#1''}%
\providecommand \bibnamefont  [1]{#1}%
\providecommand \bibfnamefont [1]{#1}%
\providecommand \citenamefont [1]{#1}%
\providecommand \href@noop [0]{\@secondoftwo}%
\providecommand \href[0]{\begingroup \@sanitize@url \@href}%
\providecommand \@href[1]{\@@startlink{#1}\@@href}%
\providecommand \@@href[1]{\endgroup#1\@@endlink}%
\providecommand \@sanitize@url [0]{\catcode `\\12\catcode `\$12\catcode
  `\&12\catcode `\#12\catcode `\^12\catcode `\_12\catcode `\%12\relax}%
\providecommand \@@startlink[1]{}%
\providecommand \@@endlink[0]{}%
\providecommand \url  [0]{\begingroup\@sanitize@url \@url }%
\providecommand \@url [1]{\endgroup\@href {#1}{\urlprefix }}%
\providecommand \urlprefix  [0]{URL }%
\providecommand \Eprint [0]{\href }%
\providecommand \doibase [0]{http://dx.doi.org/}%
\providecommand \selectlanguage [0]{\@gobble}%
\providecommand \bibinfo  [0]{\@secondoftwo}%
\providecommand \bibfield  [0]{\@secondoftwo}%
\providecommand \translation [1]{[#1]}%
\providecommand \BibitemOpen [0]{}%
\providecommand \bibitemStop [0]{}%
\providecommand \bibitemNoStop [0]{.\EOS\space}%
\providecommand \EOS [0]{\spacefactor3000\relax}%
\providecommand \BibitemShut  [1]{\csname bibitem#1\endcsname}%
\let\auto@bib@innerbib\@empty
\bibitem [{\citenamefont {Desjarlais}\ \emph {et~al.}(2002)\citenamefont
  {Desjarlais}, \citenamefont {Kress},\ and\ \citenamefont
  {Collins}}]{Desjarlais02}%
  \BibitemOpen
  \bibfield  {author} {\bibinfo {author} {\bibfnamefont {M.~P.}\ \bibnamefont
  {Desjarlais}}, \bibinfo {author} {\bibfnamefont {J.~D.}\ \bibnamefont
  {Kress}}, \ and\ \bibinfo {author} {\bibfnamefont {L.~A.}\ \bibnamefont
  {Collins}},\ }\enquote {\bibinfo {title} {Electrical conductivity for warm,
  dense aluminum plasmas and liquids},}\ \href{\doibase
  10.1103/PhysRevE.66.025401} {\bibfield  {journal} {\bibinfo  {journal} {Phys.
  Rev. E}\ }\textbf {\bibinfo {volume} {66}},\ \bibinfo {pages} {025401}
  (\bibinfo {year} {2002})}\BibitemShut {NoStop}%
\bibitem [{\citenamefont {Chen}\ \emph {et~al.}(2013)\citenamefont {Chen},
  \citenamefont {Holst}, \citenamefont {Kirkwood}, \citenamefont {Sametoglu},
  \citenamefont {Reid}, \citenamefont {Tsui}, \citenamefont {Recoules},\ and\
  \citenamefont {Ng}}]{Chen13}%
  \BibitemOpen
  \bibfield  {author} {\bibinfo {author} {\bibfnamefont {Z.}~\bibnamefont
  {Chen}}, \bibinfo {author} {\bibfnamefont {B.}~\bibnamefont {Holst}},
  \bibinfo {author} {\bibfnamefont {S.~E.}\ \bibnamefont {Kirkwood}}, \bibinfo
  {author} {\bibfnamefont {V.}~\bibnamefont {Sametoglu}},  \emph {et~al.},\
  }\enquote {\bibinfo {title} {Evolution of ac Conductivity in Nonequilibrium
  Warm Dense Gold},}\ \href{\doibase 10.1103/PhysRevLett.110.135001} {\bibfield
   {journal} {\bibinfo  {journal} {Phys. Rev. Lett.}\ }\textbf {\bibinfo
  {volume} {110}},\ \bibinfo {pages} {135001} (\bibinfo {year}
  {2013})}\BibitemShut {NoStop}%
\bibitem [{\citenamefont {Elsayed-Ali}\ \emph {et~al.}(1987)\citenamefont
  {Elsayed-Ali}, \citenamefont {Norris}, \citenamefont {Pessot},\ and\
  \citenamefont {Mourou}}]{Elsayed-Ali87}%
  \BibitemOpen
  \bibfield  {author} {\bibinfo {author} {\bibfnamefont {H.~E.}\ \bibnamefont
  {Elsayed-Ali}}, \bibinfo {author} {\bibfnamefont {T.~B.}\ \bibnamefont
  {Norris}}, \bibinfo {author} {\bibfnamefont {M.~A.}\ \bibnamefont {Pessot}},
  \ and\ \bibinfo {author} {\bibfnamefont {G.~A.}\ \bibnamefont {Mourou}},\
  }\enquote {\bibinfo {title} {Time-resolved observation of electron-phonon
  relaxation in copper},}\ \href{\doibase 10.1103/PhysRevLett.58.1212}
  {\bibfield  {journal} {\bibinfo  {journal} {Phys. Rev. Lett.}\ }\textbf
  {\bibinfo {volume} {58}},\ \bibinfo {pages} {1212} (\bibinfo {year}
  {1987})}\BibitemShut {NoStop}%
\bibitem [{\citenamefont {Schoenlein}\ \emph {et~al.}(1987)\citenamefont
  {Schoenlein}, \citenamefont {Lin}, \citenamefont {Fujimoto},\ and\
  \citenamefont {Eesley}}]{Schoenlein87}%
  \BibitemOpen
  \bibfield  {author} {\bibinfo {author} {\bibfnamefont {R.~W.}\ \bibnamefont
  {Schoenlein}}, \bibinfo {author} {\bibfnamefont {W.~Z.}\ \bibnamefont {Lin}},
  \bibinfo {author} {\bibfnamefont {J.~G.}\ \bibnamefont {Fujimoto}}, \ and\
  \bibinfo {author} {\bibfnamefont {G.~L.}\ \bibnamefont {Eesley}},\ }\enquote
  {\bibinfo {title} {Femtosecond studies of nonequilibrium electronic processes
  in metals},}\ \href{\doibase 10.1103/PhysRevLett.58.1680} {\bibfield
  {journal} {\bibinfo  {journal} {Phys. Rev. Lett.}\ }\textbf {\bibinfo
  {volume} {58}},\ \bibinfo {pages} {1680} (\bibinfo {year}
  {1987})}\BibitemShut {NoStop}%
\bibitem [{\citenamefont {Recoules}\ \emph {et~al.}(2006)\citenamefont
  {Recoules}, \citenamefont {Cl\'erouin}, \citenamefont {Z\'erah},
  \citenamefont {Anglade},\ and\ \citenamefont {Mazevet}}]{Recoules06}%
  \BibitemOpen
  \bibfield  {author} {\bibinfo {author} {\bibfnamefont {V.}~\bibnamefont
  {Recoules}}, \bibinfo {author} {\bibfnamefont {J.}~\bibnamefont
  {Cl\'erouin}}, \bibinfo {author} {\bibfnamefont {G.}~\bibnamefont {Z\'erah}},
  \bibinfo {author} {\bibfnamefont {P.~M.}\ \bibnamefont {Anglade}}, \ and\
  \bibinfo {author} {\bibfnamefont {S.}~\bibnamefont {Mazevet}},\ }\enquote
  {\bibinfo {title} {Effect of Intense Laser Irradiation on the Lattice
  Stability of Semiconductors and Metals},}\ \href{\doibase
  10.1103/PhysRevLett.96.055503} {\bibfield  {journal} {\bibinfo  {journal}
  {Phys. Rev. Lett.}\ }\textbf {\bibinfo {volume} {96}},\ \bibinfo {pages}
  {055503} (\bibinfo {year} {2006})}\BibitemShut {NoStop}%
\bibitem [{\citenamefont {Lin}\ \emph {et~al.}(2008)\citenamefont {Lin},
  \citenamefont {Zhigilei},\ and\ \citenamefont {Celli}}]{Lin08}%
  \BibitemOpen
  \bibfield  {author} {\bibinfo {author} {\bibfnamefont {Z.}~\bibnamefont
  {Lin}}, \bibinfo {author} {\bibfnamefont {L.~V.}\ \bibnamefont {Zhigilei}}, \
  and\ \bibinfo {author} {\bibfnamefont {V.}~\bibnamefont {Celli}},\ }\enquote
  {\bibinfo {title} {Electron-phonon coupling and electron heat capacity of
  metals under conditions of strong electron-phonon nonequilibrium},}\
  \href{\doibase 10.1103/PhysRevB.77.075133} {\bibfield  {journal} {\bibinfo
  {journal} {Phys. Rev. B}\ }\textbf {\bibinfo {volume} {77}},\ \bibinfo
  {pages} {075133} (\bibinfo {year} {2008})}\BibitemShut {NoStop}%
\bibitem [{\citenamefont {Petrov}\ \emph {et~al.}(2013)\citenamefont {Petrov},
  \citenamefont {Inogamov},\ and\ \citenamefont {Migdal}}]{Petrov13}%
  \BibitemOpen
  \bibfield  {author} {\bibinfo {author} {\bibfnamefont {Y.~V.}\ \bibnamefont
  {Petrov}}, \bibinfo {author} {\bibfnamefont {N.}~\bibnamefont {Inogamov}}, \
  and\ \bibinfo {author} {\bibfnamefont {K.}~\bibnamefont {Migdal}},\ }\enquote
  {\bibinfo {title} {Thermal conductivity and the electron-ion heat transfer
  coefficient in condensed media with a strongly excited electron subsystem},}\
  \href@noop {} {\bibfield  {journal} {\bibinfo  {journal} {JETP letters}\
  }\textbf {\bibinfo {volume} {97}},\ \bibinfo {pages} {20} (\bibinfo {year}
  {2013})}\BibitemShut {NoStop}%
\bibitem [{\citenamefont {Mueller}\ and\ \citenamefont
  {Rethfeld}(2013)}]{Mueller13}%
  \BibitemOpen
  \bibfield  {author} {\bibinfo {author} {\bibfnamefont {B.~Y.}\ \bibnamefont
  {Mueller}}\ and\ \bibinfo {author} {\bibfnamefont {B.}~\bibnamefont
  {Rethfeld}},\ }\enquote {\bibinfo {title} {Relaxation dynamics in
  laser-excited metals under nonequilibrium conditions},}\ \href{\doibase
  10.1103/PhysRevB.87.035139} {\bibfield  {journal} {\bibinfo  {journal} {Phys.
  Rev. B}\ }\textbf {\bibinfo {volume} {87}},\ \bibinfo {pages} {035139}
  (\bibinfo {year} {2013})}\BibitemShut {NoStop}%
\bibitem [{\citenamefont {B\'evillon}\ \emph {et~al.}(2014)\citenamefont
  {B\'evillon}, \citenamefont {Colombier}, \citenamefont {Recoules},\ and\
  \citenamefont {Stoian}}]{Bevillon14a}%
  \BibitemOpen
  \bibfield  {author} {\bibinfo {author} {\bibfnamefont {E.}~\bibnamefont
  {B\'evillon}}, \bibinfo {author} {\bibfnamefont {J.~P.}\ \bibnamefont
  {Colombier}}, \bibinfo {author} {\bibfnamefont {V.}~\bibnamefont {Recoules}},
  \ and\ \bibinfo {author} {\bibfnamefont {R.}~\bibnamefont {Stoian}},\
  }\enquote {\bibinfo {title} {Free-electron properties of metals under
  ultrafast laser-induced electron-phonon nonequilibrium: A first-principles
  study},}\ \href{\doibase 10.1103/PhysRevB.89.115117} {\bibfield  {journal}
  {\bibinfo  {journal} {Phys. Rev. B}\ }\textbf {\bibinfo {volume} {89}},\
  \bibinfo {pages} {115117} (\bibinfo {year} {2014})}\BibitemShut {NoStop}%
\bibitem [{\citenamefont {MacDonald}\ \emph {et~al.}(2009)\citenamefont
  {MacDonald}, \citenamefont {S\'{a}mson}, \citenamefont {Stockman},\ and\
  \citenamefont {Zheludev}}]{MacDonald09}%
  \BibitemOpen
  \bibfield  {author} {\bibinfo {author} {\bibfnamefont {K.~F.}\ \bibnamefont
  {MacDonald}}, \bibinfo {author} {\bibfnamefont {Z.~L.}\ \bibnamefont
  {S\'{a}mson}}, \bibinfo {author} {\bibfnamefont {M.~I.}\ \bibnamefont
  {Stockman}}, \ and\ \bibinfo {author} {\bibfnamefont {N.~I.}\ \bibnamefont
  {Zheludev}},\ }\enquote {\bibinfo {title} {Ultrafast active plasmonics},}\
  \href{\doibase 10.1038/nphoton.2008.249} {\bibfield  {journal} {\bibinfo
  {journal} {Nat. Photon.}\ }\textbf {\bibinfo {volume} {3}},\ \bibinfo {pages}
  {55} (\bibinfo {year} {2009})}\BibitemShut {NoStop}%
\bibitem [{\citenamefont {Luk'yanchuk}\ \emph {et~al.}(2010)\citenamefont
  {Luk'yanchuk}, \citenamefont {Zheludev}, \citenamefont {Maier}, \citenamefont
  {Halas}, \citenamefont {Nordlander}, \citenamefont {Giessen},\ and\
  \citenamefont {Chong}}]{Lukyanchuk10}%
  \BibitemOpen
  \bibfield  {author} {\bibinfo {author} {\bibfnamefont {B.}~\bibnamefont
  {Luk'yanchuk}}, \bibinfo {author} {\bibfnamefont {N.~I.}\ \bibnamefont
  {Zheludev}}, \bibinfo {author} {\bibfnamefont {S.~A.}\ \bibnamefont {Maier}},
  \bibinfo {author} {\bibfnamefont {N.~J.}\ \bibnamefont {Halas}}, \bibinfo
  {author} {\bibfnamefont {P.}~\bibnamefont {Nordlander}}, \bibinfo {author}
  {\bibfnamefont {H.}~\bibnamefont {Giessen}}, \ and\ \bibinfo {author}
  {\bibfnamefont {C.~T.}\ \bibnamefont {Chong}},\ }\enquote {\bibinfo {title}
  {The Fano resonance in plasmonic nanostructures and metamaterials},}\
  \href{\doibase 10.1038/nmat2810} {\bibfield  {journal} {\bibinfo  {journal}
  {Nat. Mater.}\ }\textbf {\bibinfo {volume} {9}},\ \bibinfo {pages} {707}
  (\bibinfo {year} {2010})}\BibitemShut {NoStop}%
\bibitem [{\citenamefont {Hessel}\ and\ \citenamefont
  {Oliner}(1965)}]{Hessel65}%
  \BibitemOpen
  \bibfield  {author} {\bibinfo {author} {\bibfnamefont {A.}~\bibnamefont
  {Hessel}}\ and\ \bibinfo {author} {\bibfnamefont {A.~A.}\ \bibnamefont
  {Oliner}},\ }\enquote {\bibinfo {title} {A New Theory of Wood's Anomalies on
  Optical Gratings},}\ \href{\doibase 10.1364/AO.4.001275} {\bibfield
  {journal} {\bibinfo  {journal} {Appl. Opt.}\ }\textbf {\bibinfo {volume}
  {4}},\ \bibinfo {pages} {1275} (\bibinfo {year} {1965})}\BibitemShut
  {NoStop}%
\bibitem [{\citenamefont {Birnbaum}(1965)}]{Birnbaum65}%
  \BibitemOpen
  \bibfield  {author} {\bibinfo {author} {\bibfnamefont {M.}~\bibnamefont
  {Birnbaum}},\ }\enquote {\bibinfo {title} {Semiconductor Surface Damage
  Produced by Ruby Lasers},}\ \href{\doibase 10.1063/1.1703071} {\bibfield
  {journal} {\bibinfo  {journal} {J. Appl. Phys.}\ }\textbf {\bibinfo {volume}
  {36}},\ \bibinfo {pages} {3688} (\bibinfo {year} {1965})}\BibitemShut
  {NoStop}%
\bibitem [{\citenamefont {Zorba}\ \emph {et~al.}(2008)\citenamefont {Zorba},
  \citenamefont {Stratakis}, \citenamefont {Barberoglou}, \citenamefont
  {Spanakis}, \citenamefont {Tzanetakis}, \citenamefont {Anastasiadis},\ and\
  \citenamefont {Fotakis}}]{Zorba08}%
  \BibitemOpen
  \bibfield  {author} {\bibinfo {author} {\bibfnamefont {V.}~\bibnamefont
  {Zorba}}, \bibinfo {author} {\bibfnamefont {E.}~\bibnamefont {Stratakis}},
  \bibinfo {author} {\bibfnamefont {M.}~\bibnamefont {Barberoglou}}, \bibinfo
  {author} {\bibfnamefont {E.}~\bibnamefont {Spanakis}}, \bibinfo {author}
  {\bibfnamefont {P.}~\bibnamefont {Tzanetakis}}, \bibinfo {author}
  {\bibfnamefont {S.~H.}\ \bibnamefont {Anastasiadis}}, \ and\ \bibinfo
  {author} {\bibfnamefont {C.}~\bibnamefont {Fotakis}},\ }\enquote {\bibinfo
  {title} {Biomimetic Artificial Surfaces Quantitatively Reproduce the Water
  Repellency of a Lotus Leaf},}\ \href{\doibase 10.1002/adma.200800651}
  {\bibfield  {journal} {\bibinfo  {journal} {Adv. Mater.}\ }\textbf {\bibinfo
  {volume} {20}},\ \bibinfo {pages} {4049} (\bibinfo {year}
  {2008})}\BibitemShut {NoStop}%
\bibitem [{\citenamefont {Dusser}\ \emph {et~al.}(2010)\citenamefont {Dusser},
  \citenamefont {Sagan}, \citenamefont {Soder}, \citenamefont {Faure},
  \citenamefont {Colombier}, \citenamefont {Jourlin},\ and\ \citenamefont
  {Audouard}}]{Dusser10}%
  \BibitemOpen
  \bibfield  {author} {\bibinfo {author} {\bibfnamefont {B.}~\bibnamefont
  {Dusser}}, \bibinfo {author} {\bibfnamefont {Z.}~\bibnamefont {Sagan}},
  \bibinfo {author} {\bibfnamefont {H.}~\bibnamefont {Soder}}, \bibinfo
  {author} {\bibfnamefont {N.}~\bibnamefont {Faure}}, \bibinfo {author}
  {\bibfnamefont {J.-P.}\ \bibnamefont {Colombier}}, \bibinfo {author}
  {\bibfnamefont {M.}~\bibnamefont {Jourlin}}, \ and\ \bibinfo {author}
  {\bibfnamefont {E.}~\bibnamefont {Audouard}},\ }\enquote {\bibinfo {title}
  {Controlled nanostructrures formation by ultra fast laser pulses for color
  marking},}\ \href@noop {} {\bibfield  {journal} {\bibinfo  {journal} {Optics
  express}\ }\textbf {\bibinfo {volume} {18}},\ \bibinfo {pages} {2913}
  (\bibinfo {year} {2010})}\BibitemShut {NoStop}%
\bibitem [{\citenamefont {Zhang}\ \emph {et~al.}(2014)\citenamefont {Zhang},
  \citenamefont {Gecevi\ifmmode~\check{c}\else \v{c}\fi{}ius}, \citenamefont
  {Beresna},\ and\ \citenamefont {Kazansky}}]{Zhang14}%
  \BibitemOpen
  \bibfield  {author} {\bibinfo {author} {\bibfnamefont {J.}~\bibnamefont
  {Zhang}}, \bibinfo {author} {\bibfnamefont {M.}~\bibnamefont
  {Gecevi\ifmmode~\check{c}\else \v{c}\fi{}ius}}, \bibinfo {author}
  {\bibfnamefont {M.}~\bibnamefont {Beresna}}, \ and\ \bibinfo {author}
  {\bibfnamefont {P.~G.}\ \bibnamefont {Kazansky}},\ }\enquote {\bibinfo
  {title} {Seemingly Unlimited Lifetime Data Storage in Nanostructured
  Glass},}\ \href{\doibase 10.1103/PhysRevLett.112.033901} {\bibfield
  {journal} {\bibinfo  {journal} {Phys. Rev. Lett.}\ }\textbf {\bibinfo
  {volume} {112}},\ \bibinfo {pages} {033901} (\bibinfo {year}
  {2014})}\BibitemShut {NoStop}%
\bibitem [{\citenamefont {\"{O}ktem}\ \emph {et~al.}(2013)\citenamefont
  {\"{O}ktem}, \citenamefont {Pavlov}, \citenamefont {Ilday}, \citenamefont
  {Kalayc{\i}o\u{g}lu}, \citenamefont {Rybak}, \citenamefont {Yava\c{s}},
  \citenamefont {Erdo\u{g}an},\ and\ \citenamefont {Ilday}}]{Oktem13}%
  \BibitemOpen
  \bibfield  {author} {\bibinfo {author} {\bibfnamefont {B.}~\bibnamefont
  {\"{O}ktem}}, \bibinfo {author} {\bibfnamefont {I.}~\bibnamefont {Pavlov}},
  \bibinfo {author} {\bibfnamefont {S.}~\bibnamefont {Ilday}}, \bibinfo
  {author} {\bibfnamefont {H.}~\bibnamefont {Kalayc{\i}o\u{g}lu}},  \emph
  {et~al.},\ }\enquote {\bibinfo {title} {Nonlinear laser lithography for
  indefinitely large-area nanostructuring with femtosecond pulses},}\
  \href{\doibase 10.1038/nphoton.2013.272} {\bibfield  {journal} {\bibinfo
  {journal} {Nat. Photon.}\ }\textbf {\bibinfo {volume} {7}},\ \bibinfo {pages}
  {897} (\bibinfo {year} {2013})}\BibitemShut {NoStop}%
\bibitem [{\citenamefont {Sipe}\ \emph {et~al.}(1983)\citenamefont {Sipe},
  \citenamefont {Young}, \citenamefont {Preston},\ and\ \citenamefont {van
  Driel}}]{Sipe83}%
  \BibitemOpen
  \bibfield  {author} {\bibinfo {author} {\bibfnamefont {J.~E.}\ \bibnamefont
  {Sipe}}, \bibinfo {author} {\bibfnamefont {J.~F.}\ \bibnamefont {Young}},
  \bibinfo {author} {\bibfnamefont {J.~S.}\ \bibnamefont {Preston}}, \ and\
  \bibinfo {author} {\bibfnamefont {H.~M.}\ \bibnamefont {van Driel}},\
  }\enquote {\bibinfo {title} {Laser-induced periodic surface structure. I.
  Theory},}\ \href{\doibase 10.1103/PhysRevB.27.1141} {\bibfield  {journal}
  {\bibinfo  {journal} {Phys. Rev. B}\ }\textbf {\bibinfo {volume} {27}},\
  \bibinfo {pages} {1141} (\bibinfo {year} {1983})}\BibitemShut {NoStop}%
\bibitem [{\citenamefont {Vorobyev}\ and\ \citenamefont
  {Guo}(2008)}]{Vorobyev08}%
  \BibitemOpen
  \bibfield  {author} {\bibinfo {author} {\bibfnamefont {A.~Y.}\ \bibnamefont
  {Vorobyev}}\ and\ \bibinfo {author} {\bibfnamefont {C.}~\bibnamefont {Guo}},\
  }\enquote {\bibinfo {title} {Femtosecond laser-induced periodic surface
  structure formation on tungsten},}\ \href{\doibase
  http://dx.doi.org/10.1063/1.2981072} {\bibfield  {journal} {\bibinfo
  {journal} {Journal of Applied Physics}\ }\textbf {\bibinfo {volume} {104}},\
  \bibinfo {eid} {063523} (\bibinfo {year} {2008})}\BibitemShut {NoStop}%
\bibitem [{\citenamefont {Colombier}\ \emph {et~al.}(2012)\citenamefont
  {Colombier}, \citenamefont {Combis}, \citenamefont {Audouard},\ and\
  \citenamefont {Stoian}}]{Colombier12b}%
  \BibitemOpen
  \bibfield  {author} {\bibinfo {author} {\bibfnamefont {J.-P.}\ \bibnamefont
  {Colombier}}, \bibinfo {author} {\bibfnamefont {P.}~\bibnamefont {Combis}},
  \bibinfo {author} {\bibfnamefont {E.}~\bibnamefont {Audouard}}, \ and\
  \bibinfo {author} {\bibfnamefont {R.}~\bibnamefont {Stoian}},\ }\enquote
  {\bibinfo {title} {Guiding heat in laser ablation of metals on ultrafast
  timescales: an adaptive modeling approach on aluminum},}\ \href@noop {}
  {\bibfield  {journal} {\bibinfo  {journal} {New Journal of Physics}\ }\textbf
  {\bibinfo {volume} {14}},\ \bibinfo {pages} {013039} (\bibinfo {year}
  {2012})}\BibitemShut {NoStop}%
\bibitem [{\citenamefont {Gonze}\ \emph {et~al.}(2009)\citenamefont {Gonze},
  \citenamefont {Amadon}, \citenamefont {Anglade}, \citenamefont {Beuken},
  \citenamefont {Bottin}, \citenamefont {Boulanger}, \citenamefont {Bruneval},
  \citenamefont {Caliste}, \citenamefont {Caracas}, \citenamefont {Cote} \emph
  {et~al.}}]{Gonze09}%
  \BibitemOpen
  \bibfield  {author} {\bibinfo {author} {\bibfnamefont {X.}~\bibnamefont
  {Gonze}}, \bibinfo {author} {\bibfnamefont {B.}~\bibnamefont {Amadon}},
  \bibinfo {author} {\bibfnamefont {P.-M.}\ \bibnamefont {Anglade}}, \bibinfo
  {author} {\bibfnamefont {J.-M.}\ \bibnamefont {Beuken}},  \emph {et~al.},\
  }\enquote {\bibinfo {title} {ABINIT: First-principles approach to material
  and nanosystem properties},}\ \href@noop {} {\bibfield  {journal} {\bibinfo
  {journal} {Computer Physics Communications}\ }\textbf {\bibinfo {volume}
  {180}},\ \bibinfo {pages} {2582} (\bibinfo {year} {2009})}\BibitemShut
  {NoStop}%
\bibitem [{\citenamefont {Hohenberg}\ and\ \citenamefont
  {Kohn}(1964)}]{Hohenberg64}%
  \BibitemOpen
  \bibfield  {author} {\bibinfo {author} {\bibfnamefont {P.}~\bibnamefont
  {Hohenberg}}\ and\ \bibinfo {author} {\bibfnamefont {W.}~\bibnamefont
  {Kohn}},\ }\enquote {\bibinfo {title} {Inhomogeneous Electron Gas},}\
  \href{\doibase 10.1103/PhysRev.136.B864} {\bibfield  {journal} {\bibinfo
  {journal} {Phys. Rev.}\ }\textbf {\bibinfo {volume} {136}},\ \bibinfo {pages}
  {B864} (\bibinfo {year} {1964})}\BibitemShut {NoStop}%
\bibitem [{\citenamefont {Kohn}\ and\ \citenamefont {Sham}(1965)}]{Kohn65}%
  \BibitemOpen
  \bibfield  {author} {\bibinfo {author} {\bibfnamefont {W.}~\bibnamefont
  {Kohn}}\ and\ \bibinfo {author} {\bibfnamefont {L.~J.}\ \bibnamefont
  {Sham}},\ }\enquote {\bibinfo {title} {Self-Consistent Equations Including
  Exchange and Correlation Effects},}\ \href{\doibase
  10.1103/PhysRev.140.A1133} {\bibfield  {journal} {\bibinfo  {journal} {Phys.
  Rev.}\ }\textbf {\bibinfo {volume} {140}},\ \bibinfo {pages} {A1133}
  (\bibinfo {year} {1965})}\BibitemShut {NoStop}%
\bibitem [{\citenamefont {Mermin}(1965)}]{Mermin65}%
  \BibitemOpen
  \bibfield  {author} {\bibinfo {author} {\bibfnamefont {N.~D.}\ \bibnamefont
  {Mermin}},\ }\enquote {\bibinfo {title} {Thermal Properties of the
  Inhomogeneous Electron Gas},}\ \href{\doibase 10.1103/PhysRev.137.A1441}
  {\bibfield  {journal} {\bibinfo  {journal} {Phys. Rev.}\ }\textbf {\bibinfo
  {volume} {137}},\ \bibinfo {pages} {A1441} (\bibinfo {year}
  {1965})}\BibitemShut {NoStop}%
\bibitem [{\citenamefont {Perdew}\ \emph {et~al.}(1996)\citenamefont {Perdew},
  \citenamefont {Burke},\ and\ \citenamefont {Ernzerhof}}]{Perdew96}%
  \BibitemOpen
  \bibfield  {author} {\bibinfo {author} {\bibfnamefont {J.~P.}\ \bibnamefont
  {Perdew}}, \bibinfo {author} {\bibfnamefont {K.}~\bibnamefont {Burke}}, \
  and\ \bibinfo {author} {\bibfnamefont {M.}~\bibnamefont {Ernzerhof}},\
  }\enquote {\bibinfo {title} {Generalized Gradient Approximation Made
  Simple},}\ \href{\doibase 10.1103/PhysRevLett.77.3865} {\bibfield  {journal}
  {\bibinfo  {journal} {Phys. Rev. Lett.}\ }\textbf {\bibinfo {volume} {77}},\
  \bibinfo {pages} {3865} (\bibinfo {year} {1996})}\BibitemShut {NoStop}%
\bibitem [{\citenamefont {Torrent}\ \emph {et~al.}(2008)\citenamefont
  {Torrent}, \citenamefont {Jollet}, \citenamefont {Bottin}, \citenamefont
  {Z\'{e}rah},\ and\ \citenamefont {Gonze}}]{Torrent08}%
  \BibitemOpen
  \bibfield  {author} {\bibinfo {author} {\bibfnamefont {M.}~\bibnamefont
  {Torrent}}, \bibinfo {author} {\bibfnamefont {F.}~\bibnamefont {Jollet}},
  \bibinfo {author} {\bibfnamefont {F.}~\bibnamefont {Bottin}}, \bibinfo
  {author} {\bibfnamefont {G.}~\bibnamefont {Z\'{e}rah}}, \ and\ \bibinfo
  {author} {\bibfnamefont {X.}~\bibnamefont {Gonze}},\ }\enquote {\bibinfo
  {title} {Implementation of the projector augmented-wave method in the ABINIT
  code: Application to the study of iron under pressure},}\ \href@noop {}
  {\bibfield  {journal} {\bibinfo  {journal} {Comput. Mater. Sci.}\ }\textbf
  {\bibinfo {volume} {42}},\ \bibinfo {pages} {337} (\bibinfo {year}
  {2008})}\BibitemShut {NoStop}%
\bibitem [{\citenamefont {Mazevet}\ \emph {et~al.}(2010)\citenamefont
  {Mazevet}, \citenamefont {Torrent}, \citenamefont {Recoules},\ and\
  \citenamefont {Jollet}}]{Mazevet10}%
  \BibitemOpen
  \bibfield  {author} {\bibinfo {author} {\bibfnamefont {S.}~\bibnamefont
  {Mazevet}}, \bibinfo {author} {\bibfnamefont {M.}~\bibnamefont {Torrent}},
  \bibinfo {author} {\bibfnamefont {V.}~\bibnamefont {Recoules}}, \ and\
  \bibinfo {author} {\bibfnamefont {F.}~\bibnamefont {Jollet}},\ }\enquote
  {\bibinfo {title} {Calculations of the transport properties within the PAW
  formalism},}\ \href@noop {} {\bibfield  {journal} {\bibinfo  {journal} {High
  Energy Density Physics}\ }\textbf {\bibinfo {volume} {6}},\ \bibinfo {pages}
  {84} (\bibinfo {year} {2010})}\BibitemShut {NoStop}%
\bibitem [{\citenamefont {Weaver}\ \emph {et~al.}(1975)\citenamefont {Weaver},
  \citenamefont {Olson},\ and\ \citenamefont {Lynch}}]{Weaver75}%
  \BibitemOpen
  \bibfield  {author} {\bibinfo {author} {\bibfnamefont {J.~H.}\ \bibnamefont
  {Weaver}}, \bibinfo {author} {\bibfnamefont {C.~G.}\ \bibnamefont {Olson}}, \
  and\ \bibinfo {author} {\bibfnamefont {D.~W.}\ \bibnamefont {Lynch}},\
  }\enquote {\bibinfo {title} {Optical properties of crystalline tungsten},}\
  \href{\doibase 10.1103/PhysRevB.12.1293} {\bibfield  {journal} {\bibinfo
  {journal} {Phys. Rev. B}\ }\textbf {\bibinfo {volume} {12}},\ \bibinfo
  {pages} {1293} (\bibinfo {year} {1975})}\BibitemShut {NoStop}%
\bibitem [{\citenamefont {Romaniello}\ \emph {et~al.}(2006)\citenamefont
  {Romaniello}, \citenamefont {de~Boeij}, \citenamefont {Carbone},\ and\
  \citenamefont {van~der Marel}}]{Romaniello06}%
  \BibitemOpen
  \bibfield  {author} {\bibinfo {author} {\bibfnamefont {P.}~\bibnamefont
  {Romaniello}}, \bibinfo {author} {\bibfnamefont {P.~L.}\ \bibnamefont
  {de~Boeij}}, \bibinfo {author} {\bibfnamefont {F.}~\bibnamefont {Carbone}}, \
  and\ \bibinfo {author} {\bibfnamefont {D.}~\bibnamefont {van~der Marel}},\
  }\enquote {\bibinfo {title} {Optical properties of bcc transition metals in
  the range 0-40 eV},}\ \href{\doibase 10.1103/PhysRevB.73.075115} {\bibfield
  {journal} {\bibinfo  {journal} {Phys. Rev. B}\ }\textbf {\bibinfo {volume}
  {73}},\ \bibinfo {pages} {075115} (\bibinfo {year} {2006})}\BibitemShut
  {NoStop}%
\bibitem [{\citenamefont {Hopkins}\ \emph {et~al.}(2008)\citenamefont
  {Hopkins}, \citenamefont {Duda}, \citenamefont {Salaway},\ and\ \citenamefont
  {Norris}}]{Hopkins08}%
  \BibitemOpen
  \bibfield  {author} {\bibinfo {author} {\bibfnamefont {P.~E.}\ \bibnamefont
  {Hopkins}}, \bibinfo {author} {\bibfnamefont {J.~C.}\ \bibnamefont {Duda}},
  \bibinfo {author} {\bibfnamefont {J.~L.}\ \bibnamefont {Salaway},
  \bibfnamefont {R.~N. and.~Smoyer}}, \ and\ \bibinfo {author} {\bibfnamefont
  {P.~M.}\ \bibnamefont {Norris}},\ }\enquote {\bibinfo {title} {Effects of
  Intra- and Interband Transitions on Electron-phonon Coupling and Electron
  Heat Capacity after Short-pulsed Laser Heating},}\ \href{\doibase
  10.1080/15567260802591985} {\bibfield  {journal} {\bibinfo  {journal}
  {Nanosc. Microsc. Therm.}\ }\textbf {\bibinfo {volume} {12}},\ \bibinfo
  {pages} {320} (\bibinfo {year} {2008})}\BibitemShut {NoStop}%
\bibitem [{\citenamefont {Roeser}\ \emph {et~al.}(2003)\citenamefont {Roeser},
  \citenamefont {Kim}, \citenamefont {Callan}, \citenamefont {Huang},
  \citenamefont {Glezer}, \citenamefont {Siegal},\ and\ \citenamefont
  {Mazur}}]{Roes03}%
  \BibitemOpen
  \bibfield  {author} {\bibinfo {author} {\bibfnamefont {C.~A.~D.}\
  \bibnamefont {Roeser}}, \bibinfo {author} {\bibfnamefont {A.~M.-T.}\
  \bibnamefont {Kim}}, \bibinfo {author} {\bibfnamefont {J.~P.}\ \bibnamefont
  {Callan}}, \bibinfo {author} {\bibfnamefont {L.}~\bibnamefont {Huang}},
  \bibinfo {author} {\bibfnamefont {E.~N.}\ \bibnamefont {Glezer}}, \bibinfo
  {author} {\bibfnamefont {Y.}~\bibnamefont {Siegal}}, \ and\ \bibinfo {author}
  {\bibfnamefont {E.}~\bibnamefont {Mazur}},\ }\enquote {\bibinfo {title}
  {Femtosecond time-resolved dielectric function measurements by dual-angle
  reflectometry},}\ \href{\doibase DOI: 10.1063/1.1582383#} {\bibfield
  {journal} {\bibinfo  {journal} {Rev. Sci. Instrum.}\ }\textbf {\bibinfo
  {volume} {74}},\ \bibinfo {pages} {3413} (\bibinfo {year}
  {2003})}\BibitemShut {NoStop}%
\bibitem [{\citenamefont {Raki\'{c}}\ \emph {et~al.}(1998)\citenamefont
  {Raki\'{c}}, \citenamefont {Djuri\v{s}i\'{c}}, \citenamefont {Elazar},\ and\
  \citenamefont {Majewski}}]{Rakic98}%
  \BibitemOpen
  \bibfield  {author} {\bibinfo {author} {\bibfnamefont {A.~D.}\ \bibnamefont
  {Raki\'{c}}}, \bibinfo {author} {\bibfnamefont {A.~B.}\ \bibnamefont
  {Djuri\v{s}i\'{c}}}, \bibinfo {author} {\bibfnamefont {J.~M.}\ \bibnamefont
  {Elazar}}, \ and\ \bibinfo {author} {\bibfnamefont {M.~L.}\ \bibnamefont
  {Majewski}},\ }\enquote {\bibinfo {title} {Optical properties of metallic
  films for vertical-cavity optoelectronic devices},}\ \href{\doibase
  10.1364/AO.37.005271} {\bibfield  {journal} {\bibinfo  {journal} {Appl.
  Opt.}\ }\textbf {\bibinfo {volume} {37}},\ \bibinfo {pages} {5271} (\bibinfo
  {year} {1998})}\BibitemShut {NoStop}%
\bibitem [{\citenamefont {Pitarke}\ \emph {et~al.}(2007)\citenamefont
  {Pitarke}, \citenamefont {Silkin}, \citenamefont {Chulkov},\ and\
  \citenamefont {Echenique}}]{Pitarke07}%
  \BibitemOpen
  \bibfield  {author} {\bibinfo {author} {\bibfnamefont {J.~M.}\ \bibnamefont
  {Pitarke}}, \bibinfo {author} {\bibfnamefont {V.~M.}\ \bibnamefont {Silkin}},
  \bibinfo {author} {\bibfnamefont {E.~V.}\ \bibnamefont {Chulkov}}, \ and\
  \bibinfo {author} {\bibfnamefont {P.~M.}\ \bibnamefont {Echenique}},\
  }\enquote {\bibinfo {title} {Theory of surface plasmons and surface-plasmon
  polaritons},}\ \href{http://stacks.iop.org/0034-4885/70/i=1/a=R01} {\bibfield
   {journal} {\bibinfo  {journal} {Rep. Prog. Phys.}\ }\textbf {\bibinfo
  {volume} {70}},\ \bibinfo {pages} {1} (\bibinfo {year} {2007})}\BibitemShut
  {NoStop}%
\bibitem [{\citenamefont {Garrelie}\ \emph {et~al.}(2011)\citenamefont
  {Garrelie}, \citenamefont {Colombier}, \citenamefont {Pigeon}, \citenamefont
  {Tonchev}, \citenamefont {Faure}, \citenamefont {Bounhalli}, \citenamefont
  {Reynaud},\ and\ \citenamefont {Parriaux}}]{Garrelie11}%
  \BibitemOpen
  \bibfield  {author} {\bibinfo {author} {\bibfnamefont {F.}~\bibnamefont
  {Garrelie}}, \bibinfo {author} {\bibfnamefont {J.-P.}\ \bibnamefont
  {Colombier}}, \bibinfo {author} {\bibfnamefont {F.}~\bibnamefont {Pigeon}},
  \bibinfo {author} {\bibfnamefont {S.}~\bibnamefont {Tonchev}},  \emph
  {et~al.},\ }\enquote {\bibinfo {title} {Evidence of surface plasmon resonance
  in ultrafast laser-induced ripples},}\ \href{\doibase 10.1364/OE.19.009035}
  {\bibfield  {journal} {\bibinfo  {journal} {Opt. Express}\ }\textbf {\bibinfo
  {volume} {19}},\ \bibinfo {pages} {9035} (\bibinfo {year}
  {2011})}\BibitemShut {NoStop}%
\end{thebibliography}%

\end{document}